\documentclass[12pt]{article}
\usepackage{graphicx} 

\usepackage{amsmath}
\usepackage{amssymb}
\usepackage{amsthm}
\usepackage{latexsym}

\newcommand{\comment}[1]{}
\newtheorem{theorem}{Theorem}        
\newtheorem{lemma}{Lemma}[section]
\newtheorem{remark}{Remark}[section]

\begin{document}

\date{}
 
\title{{\LARGE\sf Cardy's Formula for some Dependent Percolation Models}}

\author{
{\bf Federico Camia}\\
{\small \tt federico.camia\,@\,physics.nyu.edu}\\
{\small \sl Department 
of Physics, New York University, New York, NY 10003, USA}\\
\and
{\bf Charles M.~Newman}\\
{\small \tt newman\,@\,courant.nyu.edu}\\
{\small \sl Courant Inst.~of Mathematical Sciences, 
New York University, New York, NY 10012, USA}\\
\and
{\bf Vladas Sidoravicius}\\
{\small \tt vladas\,@\,impa.br}\\
{\small \sl Instituto de Matematica Pura e Aplicada,
Rio de Janeiro, RJ, Brazil}\\
}

\maketitle

\begin{abstract} 
We prove Cardy's formula for rectangular crossing probabilities
in \emph{dependent} site percolation models that arise from a deterministic
cellular automaton with a random initial state. The cellular automaton
corresponds to the zero-temperature case of Domany's
stochastic Ising ferromagnet
on the hexagonal lattice ${\mathbb H}$ (with alternating
updates of two sublattices) \cite{domany}; 
it may also be realized on the triangular lattice
${\mathbb T}$ with flips when a site disagrees with six, five and 
sometimes four of its six neighbors. 
\end{abstract}



\section{Introduction}
It was understood by physicists since 
the early seventies that critical statistical mechanics
models should possess continuum scaling limits 
with a global conformal invariance that goes
beyond pure scale invariance.
The phenomenon is particularly 
interesting in two dimensions, where every analytic function gives 
rise to a conformal transformation 
and the local conformal transformations form an infinite
dimensional group; in that context, it 
was first studied by Belavin, Polyakov and Zamolodchikov \cite{bpz1,bpz2}.
For an introduction to
the methods of conformal field theory as 
applied to two-dimensional critical percolation, see \cite{cardy}.

Until recently, however, there was no 
rigorous mathematical proof of this phenomenon, with the
exception of the Simple Symmetric Random Walk, 
whose continuum scaling limit is Brownian Motion.
Then, S. Smirnov managed to prove \cite{smirnov, smirnov1}
existence, uniqueness and conformal
invariance of the continuum scaling limit of 
critical site percolation on the triangular lattice, obtaining
in particular conformal invariance of crossing probabilities 
and Cardy's formula for rectangular crossings \cite{cardy1,cardy}.

In this paper we show that there are some natural 
\emph{dependent} percolation models for which conformal
invariance of the crossing probabilities and Cardy's formula can be proved.
Our proof relies on Smirnov's result and on properties of the 
dependent percolation models which make them, in
a sense to be specified later, ``small perturbations'' of the independent 
model treated by Smirnov.

The dependent percolation models we consider are the distributions at time
$n \ge 1$ (including the final state as $n \to \infty$) of a discrete time
deterministic dynamical process $\sigma^n$ with state space
$\{-1,+1\}^{\mathbb L}$ consisting of assignments of $-1$ or $+1$ to a
regular lattice ${\mathbb L}$.
The initial $\sigma^0$ is
``uniformly random'', i.e., the distribution of $\sigma^0$ is a Bernoulli(1/2)
product measure.
The dynamics are those of Domany's stochastic
Ising ferromagnet \cite{domany} at zero temperature.
There are two essentially equivalent versions --- one where ${\mathbb L}$
is the hexagonal lattice ${\mathbb H}$ and one where it is the
triangular lattice ${\mathbb T}$. 
We take  ${\mathbb H}$ and ${\mathbb T}$ to be regular lattices embedded in
${\mathbb R}^2$ so that the elementary cells of ${\mathbb H}$
(resp., ${\mathbb T}$) are regular hexagons (resp., equilateral triangles).
In the first version, ${\mathbb H}$,
as a bipartite graph, is partitioned into two subsets 
${\cal A}$ and ${\cal B}$ which are alternately updated so that each
$\sigma_x$ is forced to agree with a majority of its three neighbors
(which are in the other subset). 
In the second version, all sites are updated simultaneously according to
a rule based on a deterministic pairing of the six neighbors of every site
into three pairs (see the end of Section 2 for a complete explanation).
The rule is that $\sigma_x$ flips if and only if it disagrees 
(after the previous update) with
both sites in two or more
of its three neighbor pairs; thus there is (resp., is not) a flip if the number
$D_x$ of disagreeing neighbors 
is $\ge 5$ (resp., $\le 3$) and there is also a flip for some cases of
$D_x = 4$. We note that Cardy's formula can also be verified for a modified
rule in which there is a flip if and only if $D_x \ge 5$; the case of a modified
rule where there is a flip if and only if $D_x \ge 4$ is an interesting open 
problem.

\section{Definition of the model(s) and results}
In this section we give a more detailed description of the
dependent percolation models and results.

Consider the homogeneous 
ferromagnet on the hexagonal lattice ${\mathbb H}$ with states denoted by 
$\sigma = \{ \sigma_x \}_{x \in {\mathbb H}}, \, \sigma_x = \pm 1$, 
and with (formal) Hamiltonian
\begin{equation} \label{Hamiltonian}
{\cal H} = - \sum_{ \langle x,y \rangle } \sigma_x \sigma_y ,
\end{equation}
where $\sum_{ \langle x,y \rangle }$ denotes the sum over all pairs of 
neighbor sites, each pair counted once.
The variables $\sigma_x, \sigma_y$ are called spins.
We write ${\cal N}^{\mathbb H}(x)$ for the set of 
three neighbors of $x$, and indicate with
\begin{equation} \label{variation}
\Delta_x  {\cal H} (\sigma) =
2 \sum_{y \in {\cal N}^{\mathbb H}(x)} \sigma_x \sigma_y
\end{equation}
the change in the Hamiltonian when the spin $\sigma_x$ at site $x$ is 
flipped (i.e., changes sign).

Notice that the hexagonal lattice can be partitioned 
into two subsets ${\cal A}$ and ${\cal B}$
in such a way that all three neighbors of any site
in ${\cal A}$ (resp., ${\cal B}$) are in ${\cal B}$ (resp., ${\cal A}$). 
By placing an edge between any two sites of ${\cal A}$ (resp., ${\cal B}$)
that are next-nearest neighbors in ${\mathbb H}$, the subset
${\cal A}$ (resp., ${\cal B}$) becomes a triangular lattice.
(This relation between an hexagonal lattice and its triangular
``sublattice,'' sometimes expressed in terms of a ``star-triangle
transformation,'' will be used again in Remark \ref{star-triangle} below.)
We now consider the discrete time Markov process
$\sigma^n, \, n \in {\mathbb N}$, with 
state space ${\cal S} = \{ -1,+1 \}^{\mathbb H}$,
which is the zero temperature 
limit of a model of Domany \cite{domany}, constructed as follows:
\begin{itemize}
\item The initial state $\sigma^0$ is chosen 
from a symmetric Bernoulli product measure.
\item At odd times $n = 1, 3, \dots$, 
the spins in the sublattice ${\cal A}$ are updated according to the
following rule: $\sigma_x, \, x \in {\cal A}$, 
is flipped if and only if $\Delta_x  {\cal H} (\sigma)<0$.
\item At even times $n = 2, 4, \dots$, 
the spins in the sublattice ${\cal B}$ are updated according to the
same rule as for those of the sublattice ${\cal A}$.
\end{itemize}

In order to present the main result of this paper, let us denote 
by $\sigma^{\infty}$ the final state of
the process $\sigma^n$ defined above.
$\sigma^{\infty}= \lim_{n \to \infty} \sigma^n$ 
exists with probability one, as was proved in \cite{nns},
and, like $\sigma^n$ for $1 \leq n < \infty$,
defines a dependent percolation 
model on ${\mathbb H}$. These are the the main objects of our investigation.

We will call $\delta$ the ``mesh'' of the lattice and 
consider the continuum scaling limit of the dependent percolation
model $\sigma^n$ on $\delta {\mathbb H}$ as $\delta \to 0$.
For simplicity of exposition, we will prove 
Cardy's formula in the special case of a rectangle, aligned 
with the coordinate axes and of given
cross-ratio $\eta$ (a similar approach would work 
for any domain with a ``regular'' boundary, but it would involve
dealing with more complex deformations of the boundary).
Consider a finite rectangle 
${\cal R} = {\cal R}(a,b) \equiv 
(-a/2,a/2) \times (-b/2,b/2) \subset {\mathbb R}^2$ 
with sides of
lengths $a$ and $b$, 
such that the cross-ratio $a/b$ is $\eta$.
We say that there is (in $\sigma^n$) a vertical plus-crossing 
if ${\cal R} \cap \delta {\mathbb H}$ 
contains a path of $+1$ spins from $\sigma^n$ joining the top and
bottom sides of the rectangle ${\cal R}$, 
and call $P_{\delta}(\eta; n)$ the probability of such a plus-crossing
at time $n$.
More precisely, there is a vertical plus crossing if there is a path
$x_0,x_1,\dots,x_m,x_{m+1}$ in ${\mathbb H}$ with $\sigma^n_{x_j} =+1$
for all $j$, with $\delta x_1,\dots,\delta x_m$ 
all in ${\cal R}$, and with the
line segments $\overline{\delta x_0,\delta x_1}$ and 
$\overline{\delta x_m,\delta x_{m+1}}$ 
touching respectively the top side
$[-a/2,a/2] \times \{b/2\}$ and the bottom side $[-a/2,a/2] \times \{-b/2\}$.
In the next section we will prove the following result:
\begin{theorem} \label{cardy}
For all $n \geq 1$ (including $n = \infty$),
the limit $P(\eta; n) = \lim_{\delta \to 0} P_{\delta}(\eta; n)$ 
exists and is given by Cardy's formula:
\begin{equation}
P(\eta; n) = \, F_C(\eta) \, \equiv
\frac{ \Gamma (\frac{2}{3}) }{ \Gamma (\frac{4}{3}) \Gamma (\frac{1}{3}) } \, \eta^{\frac{1}{3}} \,
{}_2 F_1 \left( \frac{1}{3}, \frac{2}{3}; \frac{4}{3} ; \eta \right).
\end{equation}
\end{theorem}

A stronger result than 
Theorem \ref{cardy} can be obtained, i.e., it is possible to prove existence,
uniqueness and conformal 
invariance of the continuum scaling limit, as proven by Smirnov \cite{smirnov, smirnov1}
for independent site percolation on the triangular lattice.
Such a result, though, requires more work and 
will be pursued in a future paper.
Here we just note
that the proof is based on showing that the limit 
for our dependent percolation models (on the hexagonal lattice)
coincides with that of Smirnov for
independent percolation on the triangular lattice, 
i.e., that the models belong to the same universality class.

The following observations are useful 
in understanding the behavior of the model and will help
in the proof of Theorem \ref{cardy}.
\begin{itemize}
\item The values of the spins in the 
sublattice ${\cal A}$ at time zero are irrelevant,
since at time $1$, after the 
first update, those values are uniquely determined by the
values of the spins in the sublattice ${\cal B}$.
\item Once the initial spin 
configuration in the sublattice ${\cal B}$ is chosen, the
dynamics is completely deterministic.
\item A site can 
no longer flip once it belongs to either a loop or ``barbell''
of constant sign in ${\mathbb H}$, where 
a loop means a simple loop
(with no subloops) and a 
barbell consists of two disjoint loops connected by a path
(we regard a loop as a degenerate barbell). 
\end{itemize}
We also note that, by studying the percolation 
properties of the final state $\sigma^\infty$ on the infinite lattice
${\mathbb H}$,
it can be shown that every site is in some barbell of
constant $\sigma^\infty$-sign \cite{cns}. 

The discrete time Markov process defined 
above can be considered a simplified version of a 
continuous time process 
where an independent (rate $1$) Poisson clock is assigned to each site
$x \in {\mathbb H}$, and the spin 
at site $x$ is updated (with the same rule as in
our discrete time process)
when the corresponding clock rings.
The percolation properties of the final state $\sigma^\infty$ of that process 
were studied, both rigorously and numerically, in \cite{hn};
the results there (about critical exponents rather than critical
crossing probabilities) 
strongly suggest that that dependent percolation model is
also in the same universality class as 
independent percolation.  
Similar stochastic processes on 
different types of lattices have been studied in various papers.
See, for example, \cite{cdn, fss, gns, nns, ns1, ns2, ns3} 
for models on ${\mathbb Z}^d$
and \cite{Howard} for a model on 
the homogeneous tree of degree three.
Such models are also discussed 
extensively in the physics literature, usually on ${\mathbb Z}^d$
(see, for example, \cite{domany} and \cite{lms}).
On the hexagonal lattice, the 
discrete time dynamics is the zero-temperature case of
Domany's dynamics \cite{domany}.
Numerical simulations have been done by Nienhius \cite{Nienhuis} and
rigorous results for both the continuous and discrete dynamics
have been obtained in \cite{cns}, including
a detailed analysis of the discrete time (synchronous) case.
The analysis of \cite{cns}
is at the heart of this paper, and we will refer to and heavily rely on
it for the proof of Theorem~\ref{cardy}, which is given in the next section.

There is an alternative, but 
equivalent, way of describing the discrete time
dynamics as a deterministic cellular
automaton on the triangular lattice ${\mathbb T}$ (with random initial state).
The initial state is again 
chosen by assigning value $+1$ or $-1$ independently, with equal probability,
to each site of the triangular lattice.
Given some site $\bar{x} \in {\mathbb T}$, group its six ${\mathbb T}$-neighbors 
$y$ in three disjoint pairs $\{ y_1^{\bar{x}}, y_2^{\bar{x}} \},
\{ y_3^{\bar{x}}, y_4^{\bar{x}} \}, \{ y_5^{\bar{x}}, y_6^{\bar{x}} \}$, so that
$y_1^{\bar{x}}$ and $y_2^{\bar{x}}$ are ${\mathbb T}$-neighbors,
and so on for the other two pairs.
Translate this construction to all sites $x \in {\mathbb T}$, thus producing
three pairs of sites  $\{ y_1^x, y_2^x \}, \{ y_3^x, y_4^x \}, \{ y_5^x, y_6^x \}$
associated to each site $x \in {\mathbb T}$.
(Note that this construction does not need to specify how
${\mathbb T}$ is embedded in ${\mathbb R}^2$.)
Site $x$ is updated at times $m = 1, 2, \ldots$ according to the following rule:
the spin at site $x$ is 
changed from $\sigma_x$ to $- \sigma_x$ if and only if 
at least two of its pairs of
neighbors have the same sign and this sign is $- \sigma_x$.
\begin{remark} \label{star-triangle}
This dynamics on the triangular lattice ${\mathbb T}$ is equivalent to
the alternating sublattice dynamics on the hexagonal lattice ${\mathbb H}$
when restricted to the sublattice ${\cal B}$ for even times
$n=2m$.
To see this, start with ${\mathbb T}$ and construct an
hexagonal lattice ${\mathbb H}'$ by means of a star-triangle transformation
(see, for example, p. 335 of \cite{grimmett})
such that a site is added at the center of each of the triangles
$(x, y_1^x, y_2^x), (x, y_3^x, y_4^x)$, and $(x, y_5^x, y_6^x)$.
${\mathbb H}'$ may be partitioned into two triangular sublattices ${\cal A}'$
and ${\cal B}'$ with ${\cal B}' = {\mathbb T}$.
It is now easy to see that the dynamics on ${\mathbb T}$ for $m=1,2,\ldots$
and the alternating sublattice dynamics on ${\mathbb H}'$ restricted to
${\cal B}'$ for even times $n=2m$ are the same.
\end{remark}

\bigskip

Theorem \ref{cardy} (and its generalizations) in this context 
means that, at all times $m \geq 0$, the crossing probabilities for the
states $\sigma^m$ of this cellular 
automaton on ${\mathbb T}$
have the same conformally invariant continuum scaling limit 
as that for critical independent percolation on ${\mathbb T}$,
despite the dependence induced by the cellular
automaton dynamics.

\bigskip

\section{Proof of Theorem \ref{cardy}}
In this final section of the paper we prove Theorem \ref{cardy}.
We follow the notation of \cite{cns} and start by giving some definitions.
Let us consider a loop $\gamma$ in the 
triangular sublattice ${\cal B}$, written as an ordered
sequence of sites $(y_0, y_1, \dots, y_n)$ 
with $n \geq 3$, which are distinct except that $y_n=y_0$.
For $i=1, \dots, n$, let $\zeta_i$ be the 
unique site in ${\cal A}$ that is an ${\mathbb H}$-neighbor
of both $y_{i-1}$ and $y_i$.
We call $\gamma$ an \emph{s-loop} if $\zeta_1, \dots, \zeta_n$ are all distinct.
Similarly, a (site-self avoiding) 
path $(y_0, y_1, \dots, y_n)$ in ${\cal B}$, between
$y_0$ and $y_n$, is called an \emph{s-path} if
$\zeta_1, \dots, \zeta_n$ are all distinct. 
Notice that any path in ${\cal B}$ between $y$ and $y'$
(seen as a collection of sites) contains an s-path between $y$ and $y'$.
An s-loop of constant sign is stable 
for the dynamics since at the next update of ${\cal A}$
the presence of the constant sign 
s-loop in ${\cal B}$ will produce a stable loop of that sign
in the hexagonal lattice.
Similarly an s-path of constant sign between $y$ and $y'$ will be stable 
if $y$ and $y'$ are stable --- e.g., if they each belong to an s-loop.
A triangular loop $x_1, x_2, x_3 \in {\cal B}$ with a common
${\mathbb H}$-neighbor $\zeta \in {\cal A}$ is called a \emph{star}; it is not 
an s-loop.
A triangular loop in ${\cal B}$ 
that is not a star is an s-loop and will be called an \emph{antistar},
while any loop in ${\cal B}$ 
that contains more than three sites contains an s-loop.

Before stating a lemma, that will be a main ingredient in the proof
of Theorem \ref{cardy}, we need one more definition.
For $(x,x')$ an ordered pair of neighbors in ${\cal B}$,
we define the ``partial cluster'' $C^{\cal B}_{(x,x')}$
to be the set of sites $y \in {\cal B}$ such that
there is a (site-self avoiding) path $x_0=x', x_1, \dots, x_n=y$
in ${\cal B}$ of constant sign in $\sigma^0$, with $x_1 \neq x$
and $(x_0=x',x_1,x)$ not forming a star.
Combining the stability properties of s-loops and s-paths just discussed,
we have the following lemma.
\begin{lemma} \label{starlemma}
An s-path $(y_0, \dots, y_m)$ in ${\cal B}$ of 
constant sign in $\sigma^0$ is stable (i.e., retains that
same sign in $\sigma^n$ for all $0 \leq n\leq \infty$)
if $C^{\cal B}_{(y_1,y_0)}$ and $C^{\cal B}_{(y_{m-1},y_m)}$
both contain s-loops.
\end{lemma}

\noindent {\bf Proof of Lemma \ref{starlemma}.}
The original s-path $(y_0, \dots, y_m)$ is stable because either 
$y_0$ and $y_m$ both belong to s-loops of constant sign in $\sigma^0$
or else there is a longer s-path of constant sign in $\sigma^0$, 
between some $y$ and $y'$
(with the original $(y_0, \dots, y_m)$ as a subpath), such that both 
$y$ and $y'$ belong to s-loops of constant sign in $\sigma^0$.
\fbox \\

\bigskip

With this preparation, we are now ready to start the proof
of Theorem \ref{cardy}.
What we will prove, roughly speaking,
is that, in the limit $\delta \to 0$, there exists a
vertical plus-crossing of ${\cal R}$ from $\sigma^n$ with $n\geq 1$, 
in ${\cal R} \cap \delta {\mathbb H}$,
if and only if there exists a vertical plus-crossing 
of ${\cal R}$ from $\sigma^0$ in ${\cal R} \cap \delta {\cal B}$.
Since ${\cal B}$ is a triangular 
lattice and the initial state $\sigma^0$ is chosen from a symmetric
Bernoulli product measure, this implies that the limit 
$P(\eta ; n) = \lim_{\delta \to 0} P_{\delta}(\eta ; n)$
exists for $n\geq 1$ and is the same as in the 
case of the crossing probability for independent site percolation on the
triangular lattice, thus proving the theorem.

Consider two rectangles, ${\cal R}'= {\cal R}(a',b')$ 
with $b'$ slightly larger than $b$ and $a'$ slightly smaller than $a$,
and
${\cal R}''= {\cal R}(a'',b'')$ 
with $b''$ slightly smaller than $b$ and $a''$ slightly larger than $a$.
Call $P'_{\delta} (a',b')$ the 
probability of a vertical
plus-crossing from $\sigma^0$ in ${\cal R}' \cap \delta {\cal B}$
joining the top and bottom sides of ${\cal R}'$
and $P''_{\delta} (a'',b'')$ the probability of a 
\emph{horizontal minus}-crossing from $\sigma^0$ in
${\cal R}'' \cap \delta {\cal B}$ joining the 
left and right sides of ${\cal R}''$. 
Note that a vertical plus crossing (on the triangular lattice
$\delta {\cal B}$) occurs if and only if a horizontal minus-crossing
does not occur.
Clearly, from \cite{smirnov, smirnov1} we have

\begin{equation} \label{lim1}
 P' (a',b') \equiv
\lim_{\delta \to 0} P'_{\delta} (a',b') =
F_C(a'/b'),
\end{equation}

\begin{equation} \label{lim2}
\lim_{a' \to a,\, b' \to b} P' (a',b') = 
P'(a,b) = F_C(\eta),
\end{equation}
and

\begin{equation} \label{lim3}
\lim_{a'' \to a,\, b'' \to b} \lim_{\delta \to 0} P''_{\delta} (a'',b'') = 1-F_C(\eta).
\end{equation}

Any vertical plus-crossing of ${\cal R}' \cap \delta {\cal B}$
at time $0$ yields a vertical plus-crossing by some s-path
$(y_0, \dots, y_m)$, which then yields at time $1$ a vertical
plus-crossing of ${\cal R} \cap \delta {\mathbb H}$ by a path
$(y_{k_1},\zeta_{k_1+1}, \dots,\zeta_{k_2}, y_{k_2})$, 
providing $a'<a$, $b'\geq b$ and $\delta$
is sufficiently small. (The reason we first take $b' > b$ and then let
$b' \to b$ is to handle the case of time $n>1$, as we shall see.)
Therefore, for small $\delta$,
\begin{equation} \label{upper-bound}
P_{\delta} (\eta; n=1) \geq P'_{\delta} (a',b').
\end{equation}
On the other hand, if there is a horizontal
minus-crossing of ${\cal R}'' \cap \delta {\cal B}$ at time $0$,
it produces a horizontal minus-crossing in 
${\cal R} \cap \delta {\mathbb H}$ at time $1$ (for small $\delta$)
which blocks any possible
vertical plus-crossing in ${\cal R} \cap \delta {\mathbb H}$
at that time; therefore, for small $\delta$,

\begin{equation} \label{lower-bound}
P_{\delta} (\eta; n=1) \leq 1 - P''_{\delta} (a'',b'').
\end{equation}
Letting $\delta \to 0$ and then 
$a',a'' \to a$ and $b',b'' \to b$ and using 
(\ref{lim2})-(\ref{lower-bound}), we conclude
that $P_{\delta} (\eta; n=1)$ converges to Cardy's formula,
$F_C(\eta)$, as $\delta \to 0$.

It remains to prove that the same is true for all times $n \geq 2$.
In order to do that, we 
first have to show that 
our vertical plus-crossing of ${\cal R}' \cap \delta {\mathbb H}$
by $(y_0,\zeta_1, \dots,\zeta_m, y_m)$
created at time $1$ doesn't ``shrink'' too much 
due to the effect of the dynamics, so that at all later times,
including $n = \infty$, there is a  
vertical plus-crossing of ${\cal R} \cap \delta {\mathbb H}$
by $(y_{k_1},\zeta_{k_1+1}, \dots,\zeta_{k_2}, y_{k_2})$.

To do this by extending the
bound (\ref{upper-bound}) to all $n\geq 1$, at the
cost of a correction to the right hand side that tends to zero
with $\delta$, we apply Lemma \ref{starlemma}.
Noting that each of the partial paths $(y_0, \dots, y_{k_1})$ and
$(y_{k_2},\dots,y_m)$ contains of the order of $(b'-b)/\delta$ sites,
we see that the lemma implies that it suffices to show that 
there is some $\beta>0$ and $K < \infty$ such that
for any deterministic $(x,x')$,
\begin{equation} \label{antistar}
P(|C^{\cal B}_{(x,x')}| \geq \ell \text{ and $C^{\cal B}_{(x,x')}$ 
contains no antistar}) \leq K \, e^{- \beta \ell}.
\end{equation}
To prove (\ref{antistar}), we 
partition ${\cal B}$ into disjoint antistars and denote by $\tau$
the collection of these antistars.
We do an algorithmic construction of $C^{\cal B}_{(x,x')}$ 
(as in, e.g., \cite{fn}), where the order of
checking the sign of sites is 
such that when the first site in an antistar from $\tau$ is checked
(and found to have the same sign as $x'$), 
then the other two sites in that antistar are checked next.
Without loss of generality, we assume that $\sigma^0_{x'} = +1$.
Then standard arguments show 
that the probability in (\ref{antistar}) is bounded by
$K \, (1 - (\frac{1}{2})^3)^{(\ell /3)}$. 

To similarly extend the 
bound (\ref{lower-bound}),
one proceeds in the same way, but considering horizontal
minus-crossings of ${\cal R}'' \cap \delta {\cal B}$ 
at time zero which produce horizontal minus-crossings of
${\cal R} \cap \delta {\mathbb H}$ at time 
$n \geq 1$.
Taking the limits 
$\delta \to 0 \, , \, a' \to a \, , \, b' \to b$ concludes the proof. \fbox \\

\end{document}